\documentclass[apjl]{emulateapj}

\usepackage{natbib}
\slugcomment{To appear in ApJL, 2011}

\shorttitle{A dim candidate companion to $\epsilon$ Cephei}
\shortauthors{Mawet et al.}

\begin{document}


\title{A dim candidate companion to $\epsilon$~Cephei}


\author{D. Mawet\altaffilmark{1}}
\affil{European Southern Observatory, Alonso de Cord\'ova 3107, Vitacura, Santiago, Chile}


\author{B. Mennesson, E. Serabyn, K. Stapelfeldt}
\affil{Jet Propulsion Laboratory, California Institute of Technology, 4800 Oak Grove Drive, Pasadena, CA 91109, USA}


\author{O. Absil\altaffilmark{2}}
\affil{Institut d'Astrophysique et de G\'eophysique de Li\`ege, University of Li\`ege, 17 All\'ee du 6 Ao\^ut, 4000 Sart Tilman, Belgium}


\altaffiltext{1}{Jet Propulsion Laboratory, California Institute of Technology, 4800 Oak Grove Drive, Pasadena, CA 91109, USA}
\altaffiltext{2}{Postdoctoral Researcher F.R.S.-FNRS (Belgium)}

\begin{abstract}
Using a vector vortex coronagraph behind the 1.5-m well-corrected subaperture (WCS) at Palomar, we detected a second object very close to $\epsilon$~Cephei, a $\delta$~Scuti F0 IV star. The candidate companion, $\sim 50$ times fainter than  $\epsilon$~Cephei, if physically associated, is a late-type K or early M star, and lies at an angular separation of 330 mas, or 1.1 $\lambda/D$ for the WCS, making it the smallest angle detection ever realized with a coronagraph in terms of $\lambda/D$ units. The projected separation of the putative companion is $\sim 8.6$ AU, most likely on a highly eccentric orbit. The recently detected near-infrared excess is thus likely not due to hot dust. Moreover, we also show that the previously reported {\it IRAS} 60 $\mu$m excess was due to source confusion on the galactic plane.
\end{abstract}





\section{Introduction}

High contrast imaging at small angles is a very useful tool to access inner regions around stars. While instrumental limits are constantly improved by the development of exoplanet imaging and characterization techniques, such capabilities can be used in stellar astrophysics to discover new systems or put constraint on known ones. Moreover, this small-angle capability allows a significant reduction in the size of potential space telescopes aimed at detecting and characterizing exoplanets, which is a cost-effective way of building exoplanet missions. 

We have used a 1.5-m subaperture of the 5.1-m Hale telescope at Palomar with a small inner working angle (IWA) phase-mask coronagraph to detect a new candidate stellar companion to the $\delta$~Scuti star $\epsilon$~Cephei. The result is interesting for two reasons. First, $\epsilon$~Cephei was claimed to have an infrared excess measured by {\it IRAS}, and recently detected by near-infrared ground-based interferometry. Such infrared excesses are generally attributed to dusty debris disks \citep{Absil2006}. We show here with an image that the origin of the near-infrared excess is a second object that is likely a late-type stellar companion. Moreover, we demonstrate that the previously reported 60 $\mu$m {\it IRAS} excess is attributable to an unrelated field object. Second, the potential companion was imaged at 1.1 diffraction beamwidths ($\lambda/D$, with the working wavelength $\lambda\simeq 2.16$ $\mu$m, and the telescope diameter $D=1.5$ m) from the star. Such a small angle detection was only possible because of our phase mask vector vortex coronagraph (VVC), demonstrating that such coronagraphs can effectively provide very small IWA.

\section{The $\delta$~Scuti $\epsilon$~Cephei}

$\epsilon$~Cephei is a bright ($V=4.2$, $H=3.7$, $K=3.5$) and nearby ($d=26.2 \pm 0.3$ pc, see \citet{vanLeeuwen2007}) F0 IV star with a temperature $T=7350$ K, of the $\delta$~Scuti type (Table~1). Privileged targets in asteroseismologic studies \citep{Bruntt2007}, $\delta$~Scutis are multi-periodic variable stars lying at the base of the Cepheid instability strip, where it crosses the main sequence. Their small amplitude (0.001--1 mag) variability is due to simultaneous radial and non-radial pulsation modes with periods between 0.25 and 5 hours. The spectral type of $\delta$~Scuti stars ranges from A0 to F5, with nominal mass and temperature around 2 $M_{\rm Sun}$ and 7500 K, respectively. For $\epsilon$~Cephei, \citet{Kennelly1999} derived a mass of 1.8 $\pm 0.2$ $M_{\rm Sun}$.
 
$\epsilon$~Cephei was claimed to be detected by {\it IRAS} at 12, 25, and 60 $\mu$ as PSC~22132+5647. The reported 60 $\mu$ flux density of $1.20\pm 0.08$ Jy is $>10$ times the expected photospheric value (the ratio of excess infrared luminosity divided by the total energy output from the photosphere was reported to be $\tau \simeq 1.56 \times 10^{-4}$). On this basis, \citet{Oudmaijer1992}, \citet{Moor2006}, and \citet{Rhee2007} identified the system as hosting a bright debris disk. In \citet{Rhee2007}, this excess was fitted with a single blackbody giving a putative ring of dust at 62 AU with a temperature of 65 K. Using space velocities UVW, lithium abundance and location on a HR diagram, \citet{Rhee2007} also estimated the age of $\epsilon$~Cephei to be a very uncertain $\sim 600$ Myr.

\begin{deluxetable}{lc}
\label{Table1}
\tabletypesize{\scriptsize}
\tablecaption{$\epsilon$~Cephei fundamental stellar properties.}
\tablehead{
\colhead{Properties} & \colhead{$\epsilon$~Cephei}  }
\startdata
Coordinates (hms)	&RA=22 15 02.19 dec=+57 02 36.91\\
Galactic coordinates (deg) &l=102.87 b=+00.39\\
Proper motion ($\arcsec$)	&$\delta$RA=0.47645 $\delta$dec=0.04999\\
Radial velocity (km/s)	&-0.6\\
Parallax ($\arcsec$) &0.3886\\
Spectral type	&F0 IV			  \\
$V$ mag 		&4.2 				 \\
$K$ mag 		&3.5				\\
Age 			&$\sim 600$ Myr	\\ 
Distance		&$26.2\pm 0.3$ pc			\\ 
Mass		&1.8 $M_{\rm Sun}$	\\ 
Radius		&2 $R_{\rm Sun}$		\\ 
Temperature  	&7350 K	 \enddata
\end{deluxetable}

In the course of an on-going survey for bright exozodiacal disks around main sequence stars, lead by one of us, the CHARA-FLUOR interferometer detected a $K'$-band excess for $\epsilon$~Cephei (Absil et al. 2011, in preparation). One possible interpretation of such near-infrared excesses is emission from hot dust close to the star \citep{Absil2006,Absil2008,Absil2009,Folco2007, Akeson2009}. The grain populations usually derived from such observations are quite intriguing, as they point towards very high dust replenishment rates, high cometary activity or unlikely major collisional events.
Note that due to sparse (u,v) coverage, CHARA-FLUOR  poorly constrains the nature of the excess or its location inside the FLUOR $1\arcsec$ FWHM field of view, suggesting that a high resolution image might be able to distinguish between a close companion and dust.



\section{Observation and instrumental setup}
We observed $\epsilon$~Cephei in the $K_s$ band on 2010 June 23, as part of our program of adaptive optics (AO) high contrast imaging of nearby young bright stars at Palomar. We took 50 frames of 4.238 s each, for a total of 212.4 s on the target. We used the Palomar Well-Corrected Subaperture (WCS) and the VVC, which have been both described in details earlier \citep{Serabyn2007, Serabyn2009, Mawet2010b}. 

Briefly, a clear, off-axis WCS is provided by a set of relay optics upstream of the Palomar AO system. The off-axis relay magnifies and shifts an off-axis sub-aperture pupil onto the deformable mirror, yielding roughly 10 cm actuator spacing in the selected pupil. This configuration reduces the aperture size from 5~m to 1.5~m, but allows operation in the extreme AO regime. Providing unprecedented image quality on a ground-based telescope ($K_s$-band Strehl ratios $S\geq 90\%$), this setup is ideal for use with the VVC \citep{Mawet2010a,Serabyn2010}. 

The VVC is based on a transparent phase-mask, with an azimuthal phase ramp centered on the image of the central star. Upon propagation to the Lyot plane, the central starlight is redistributed outside the geometric pupil area. There, a simple diaphragm known as a Lyot stop is used to reject the starlight, leaving any off-axis feature mostly unaltered in the process: the 50\% off-axis throughput, which is the commonly adopted definition of the IWA, is reached at 0.9 $\lambda/D$ for a second order VVC such as the one we used at Palomar \citep{Mawet2010a}.

\section{Image of a potential companion at 1.1 resolution elements}

\begin{figure*}[!t]
\includegraphics[scale=.3]{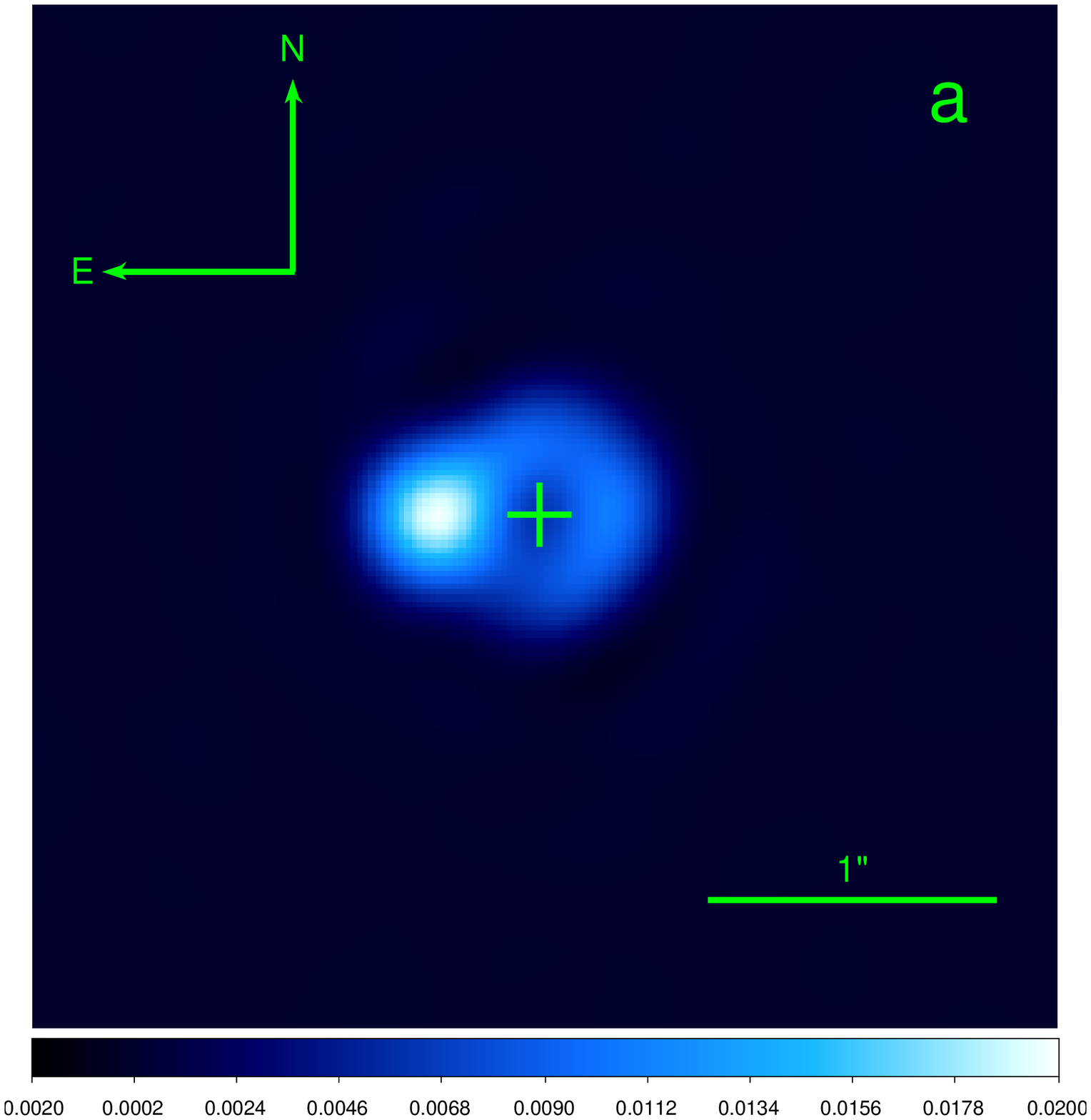}
\includegraphics[scale=.3]{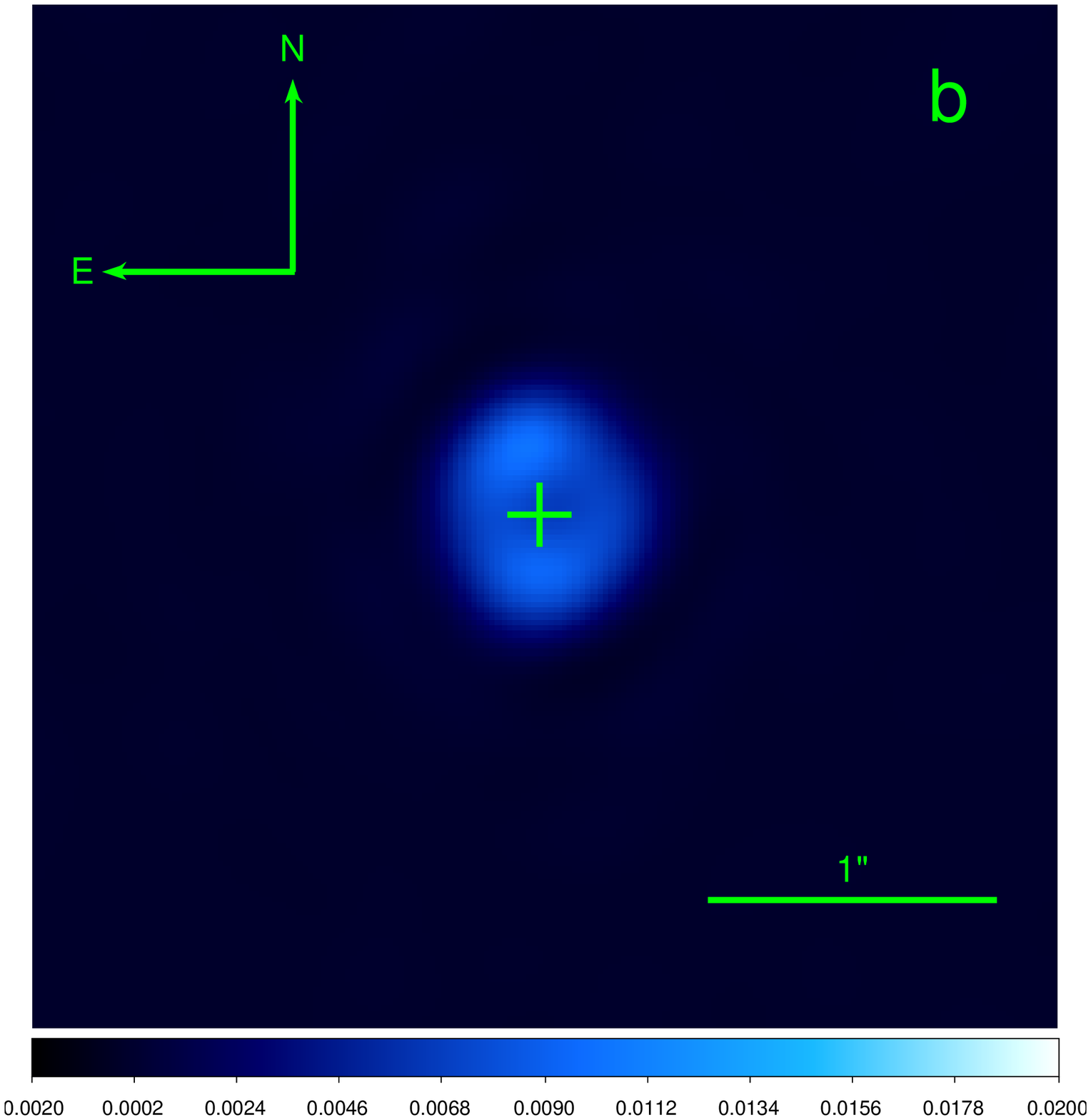}
\includegraphics[scale=.3]{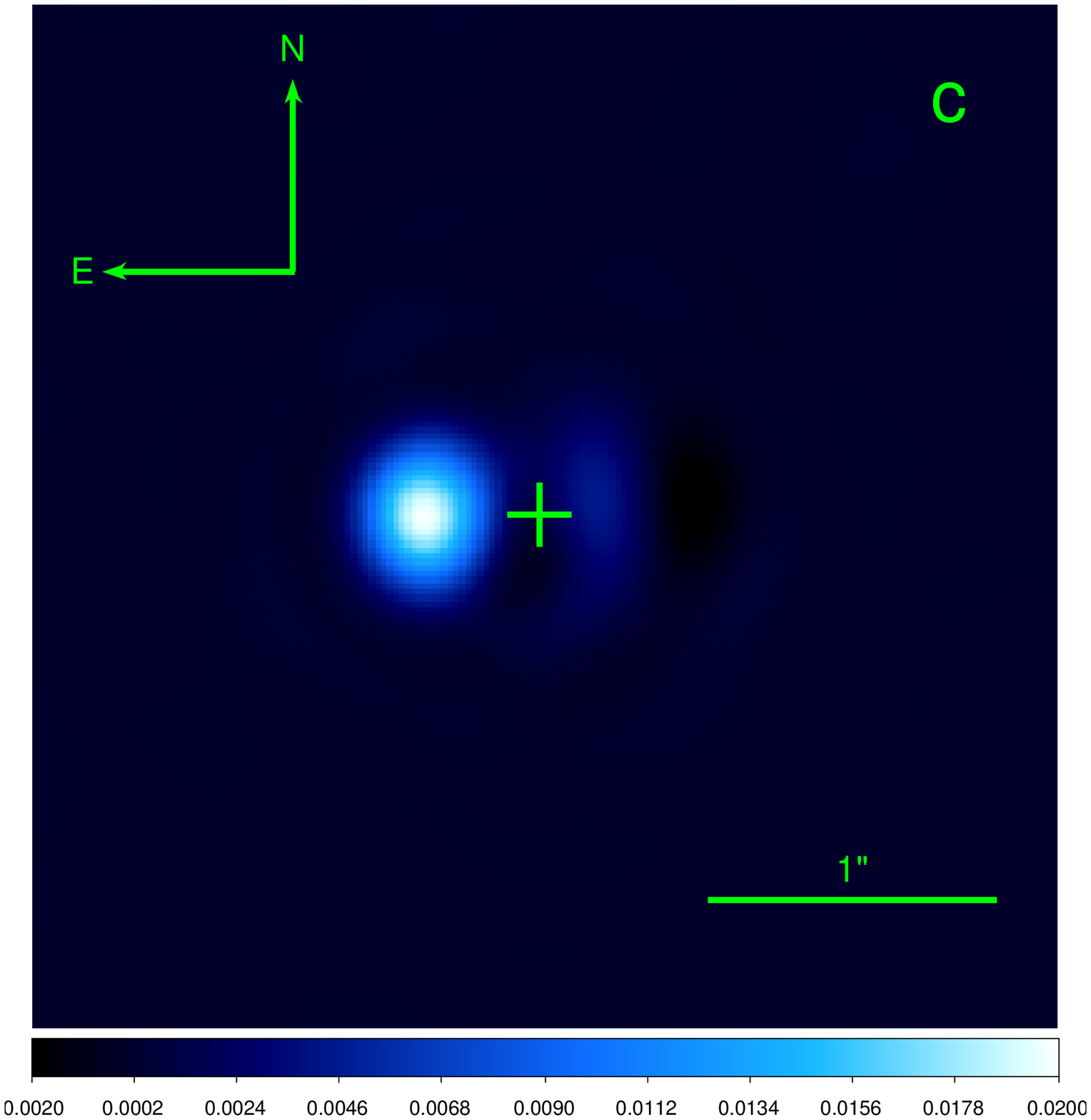}
\caption{Coronagraphic image of $\epsilon$~Cephei and its candidate companion. The scale is linear and roughly calibrated in contrast. The cross indicates the position of the star. (a): Raw image, corrected for sky background, flat field and image imperfections (bad pixels and cosmic rays). (b): Image of the reference star HD 213558. (c): Result of the subtraction of the scaled reference star image from $\epsilon$~Cephei. \label{fig1}}
\end{figure*}

The VVC allowed us to detect a second object near $\epsilon$~Cephei at 330 $\pm 50$ mas from the host star, which corresponds to $\sim 1.1$ times the width of the diffraction beam ($\lambda/D\simeq 300$ mas at $K_s$), making it the smallest angle detection ever reported with a coronagraph in $\lambda/D$ units. The projected separation of the potential companion places it at $8.6 \pm 1.4$ AU. Figure~\ref{fig1} (a) shows the raw image after sky subtraction, flat fielding, bad pixel replacement and cosmic ray removal. A faint secondary point source is clearly detected on this image to the East ($P.A.=90^\circ \pm 10^\circ$). A reference star (HD 213558) was also imaged right after $\epsilon$~Cephei, with no companion present (Figure~\ref{fig1}, b). The reference star was chosen with a $V-K$ color and brightness as similar as possible to the target. Matching $V$ magnitudes is important to ensure similar AO corrections. $K$ magnitudes, and integration time also need to be matched to ensure a proper subtraction. Also, to reduce thermal, seeing and flexure-induced variations as much as possible, the calibration star was chosen and observed as close as possible to the hour angle and elevation of the target star, and just a few minutes after the latter.

Finely scaling the reference star image to exactly match $\epsilon$~Cephei's apparent magnitude, and subtracting it from $\epsilon$~Cephei, leaves a clean detection of the second object (Figure~\ref{fig1}, c). The subtracted image was used to derive the astrometry and photometry with minimal contamination from the residual errors left after imperfect wavefront correction. Note that the $5\sigma$ detection level after subtraction is calculated to be $\sim 2 \times 10^{-3}$ at $\sim 1$ $\lambda/D$, mostly limited by residual centering errors on the VVC. 


\subsection{$K_s$-band photometry}
The 2MASS $K$ magnitude of $\epsilon$~Cephei is not very well constrained: $3.5\pm 0.3$. However, \citet{Morel1978} give an independent and accurate (few percent accuracy) measurement of the optical photometry ($B=4.47$, $V=4.19$, $R=3.92$, $I=3.77$) which is nominal for an F0 star. In this case, the $V-K$ color is 0.7 \citep{Allen4}, and the $K$ magnitude extrapolated from $V$ should be 3.49, very close to the 2MASS value.

Aperture photometry using a radius of 300 mas (our FWHM) yields a flux ratio of $0.020 \pm 0.005$ in the $K_s$ band. The apparent magnitude of the potential companion is then $K_s=7.8 \pm 0.5$. Note that we conservatively propagated the 2MASS error bars, knowing that they must be pessimistic by an order of magnitude. For an assumed distance of $26.2\pm 0.3$~pc, the absolute $K_s$ magnitude of the companion is therefore $5.70 \pm 0.52$. 

\subsection{Probability of association}\label{probability}
As yet, we have no second epoch measurement available because the Palomar AO system was taken offline in mid-2010 to enable the PALM-3000 upgrade \footnote{PALM-3000 will deliver extreme AO correction on the full 5.1-m Hale telescope \citep{Bouchez2010}.}. The probability for the potential companion to be a background object, given its proximity to $\epsilon$~Cephei, the galactic coordinates of the system (in the galactic plane but well away from the bulge, see Table 1) and the number of objects present with $K<7.8$ in a series of regions of radii 1-10 deg centered around the star (measured population in this zone of the sky), is very small, $\sim 10^{-6}$. Thus it is likely that the candidate companion is physically associated with $\epsilon$~Cephei. However, astrometry and/or spectroscopy will be needed to confirm this.

\begin{deluxetable}{lc}
\tabletypesize{\scriptsize}
\tablecaption{Candidate companion fundamental properties.}
\tablehead{
\colhead{Properties}  & \colhead{$\epsilon$~Cephei b} }
\startdata
Spectral type				&K8-M2  \\
$K$ mag 					&7.8 		\\
Age 				&$\sim 600$ Myr\\ 
Distance			&$\ge$8.6 AU from primary\\ 
Mass			&0.5 $M_{\rm Sun}$\\ 
Radius			&$\sim 0.7$ $R_{\rm Sun}$\\ 
Temperature  		&3650 K \enddata
\end{deluxetable}

\section{Discussion}
Assuming the candidate companion is bound to $\epsilon$~Cephei, we will derive its physical properties based on an evolutionary model. We will also discuss the consequence of this discovery on the interpretation of near-infrared and far-infrared excesses. An interesting question one could then ask is why this second source has largely remained undetected so far.


\subsection{Possible nature of the candidate companion}
\begin{figure}[!t]
\includegraphics[scale=.57]{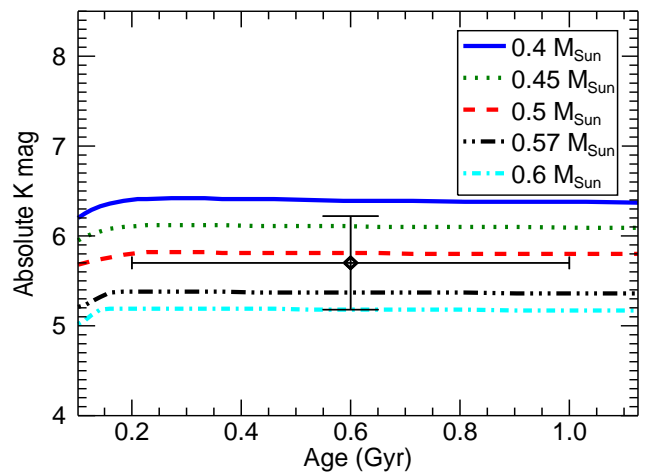}
\caption{Magnitude-age diagram for the evolutionary model from \citet{Baraffe1998}, for a range of mass 0.4 to 0.6 $M_{\rm Sun}$. Our photometry with conservative error bars is overplotted. \label{baraffe}}
\end{figure}

According to the BCAH98 evolutionary model \citep{Baraffe1998}, and following the age estimation in \citet{Rhee2007} the candidate companion would have a mass of $\sim 0.5$ $M_{\rm Sun}$, and a temperature of 3650 K (Table~2). The stellar type of the candidate companion is thus likely to be at the transition between late-type K stars and early M stars. This result is fairly independent of the age between 200 Myr up to well above 1 Gyr, making this mass determination quite robust even with big error bars (Figure~\ref{baraffe}).

The interferometrically detected near-infrared excess is likely due to the companion, and not hot dust. However, the candidate companion would only contribute $\sim 5\%$ additional flux at 60 $\mu$. It thus cannot account for the large excess detected in the {\it IRAS} beam, which was interpreted as thermal emission from cold dust at a separation of $\sim 62$ AU \citep{Rhee2007}. In this picture, the cold dust belt would thus likely be circumbinary which is not unprecedented, see for instance the case of GG Tau \citep{Krist2005}. However, as mentioned earlier, $\epsilon$~Cephei lies away from the galactic bulge but still on the galactic plane ($b=+0.4$ deg, see Table~1), which raises the possibility of source confusion given the large beam size of {\it IRAS} ($\sim 25\arcsec$ at 60 $\mu$). To confirm the excess emission, the star was observed with Spitzer/MIPS on 2009 Feb 20 (PI: George Rieke). 

\begin{figure}[!t]
\includegraphics[scale=.46]{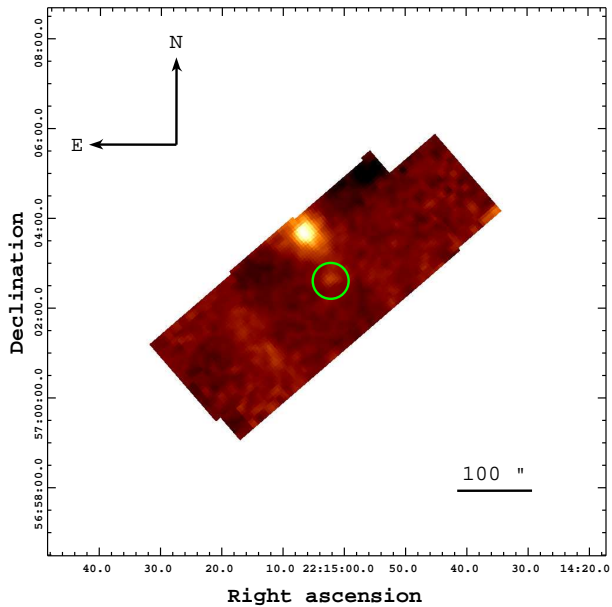}
\caption{The MIPS 70 $\mu$ BCD (Basic Calibrated Data) image of $\epsilon$~Cephei retrieved from the Spitzer Heritage Archive shows only a very weak detection at the stellar position (green circle), and a much stronger far-infrared source located $76\arcsec$ to its NNE. \label{fig2}}
\end{figure}

The MIPS 70 $\mu$ BCD (Basic Calibrated Data) image retrieved from the {\it Spitzer} Heritage Archive (see Figure~\ref{fig2}) shows only a very weak detection at the stellar position, and a much stronger far-infrared source located $76\arcsec$ to its NNE. Photometry in a $16\arcsec$ circular aperture centered on the stellar position using aperture corrections as given in the MIPS instrument handbook finds a stellar 70 $\mu$ flux density of $\sim$ 80 mJy. This value is consistent with the expected stellar photospheric emission (as extrapolated from the {\it IRAS} 12 and 25 $\mu$ flux densities). We therefore conclude that the {\it IRAS} 60 $\mu$ measurement was very likely confused by the bright adjacent source. Indeed, the reported position of the {\it IRAS} source PSC~22132+5647 is shifted $33\arcsec$ E from the known position of $\epsilon$~Cephei at the time of the measurement. Thus, it was never right to associate $\epsilon$~Cephei and PSC~22132+5647, which is likely a virtual combination of two real sources: $\epsilon$~Cephei, detected at 12 and 25 $\mu$ by {\it IRAS}, and the far-infrared source to its NNE, dominating at 60 $\mu$. In conclusion, $\epsilon$~Cephei lacks any far-infrared excess and thus does not possess a debris disk sculpted by the putative companion.

\subsection{Visible speckle interferometry}
Using speckle interferometry in the visible, \citet{HartkopfMcAlister1984}, while classifying $\epsilon$~Cephei as a spectroscopic binary (that companion would be much closer in), did not report the detection of an outer companion down to the detectivity limit of the technique used, which is $\Delta V \simeq 5$, or a flux ratio of $\sim 0.01$. 

Note that for a late K / early M type, the $V-K$ color is $\sim 3.5$ mag \citep{Allen4}. $\epsilon$~Cephei's own $V-K$ color should be only 0.7 mag \citep{Allen4}. Given our measured flux ratio of $0.020 \pm 0.005$ in the $K_s$ band, the contrast will be $(3.5-0.7)\simeq 2.8$ mag greater at $V$ band, i.e.~an overall flux ratio of $\sim 0.0015$. Thus it is not at all surprising that optical measurements (speckle interferometry or spectra) would fail to show a double-lined binary, or that {\it Hipparcos} would fail to see the companion as a separate source. 

\subsection{Absence of indirect detections}
A 0.5 $M_{\rm Sun}$ companion on a circular 8.6 AU orbit would cause an $\sim 70$ mas reflex motion of $\epsilon$~Cephei over a period of $\sim 17$ years. Such a large astrometric signature would have been easily detected by {\it Hipparcos}, which observed $\epsilon$~Cephei more than 100 times over its 3.5 year mission lifetime with a typical accuracy of 0.5 mas per observation \citep{Perryman1997}. $\epsilon$~Cephei is however classified as a single star in the {\it Hipparcos} catalog. This either suggests that our detected object is not physically associated, which seems unlikely based on a statistical argument (see Section~\ref{probability}), or that the orbital semi-major axis is actually much larger than the  apparent separation of $\sim 8.6$ AU, resulting in a much larger period than the {\it Hipparcos} mission duration. 

A first possibility is that the companion is on a significantly eccentric orbit. While no proper motion variation was observed between short-term {\it Hipparcos} and long-term Tycho observations, some proper motion acceleration was detected by {\it Hipparcos} around $\epsilon$~Cephei \citep{MakarovandKaplan2005}. According to the same authors, such behavior not only suggests that $\epsilon$~Cephei is an astrometric binary, but also that the system orbital period should be typically less than $\sim 6$ years, unless there is significant eccentricity in the system. Following this interpretation, our measured minimum orbital period of $\sim 17$ years supports a substantial eccentricity. Additionally, the system may be seen under a high inclination, resulting in a strongly elliptical apparent orbit whatever its true eccentricity. Evidence for such a significant departure from pole-on inclination comes from the observed $v\sin{i}$ of 91 km/s \citep{Royer2007}, which is to be compared with a break-up velocity around 400km/s for a sub-giant star like $\epsilon$~Cephei.

The candidate companion did not leave any detected radial velocity (RV) signature either. Indeed, \citet{Kennelly1999} report a very rich spectrum of pulsating modes and beats for $\epsilon$~Cephei, consistent with a $\delta$~Scuti (see also \citet{Gray1971}). The stellar pulsations induce RV signatures at the $\sim10$ km/s level, making it very difficult to detect faint companions using RV measurements. Indeed, assuming a semi-major axis $a \simeq 8.6$ AU, $M_1=1.8$ $M_{\rm Sun}$, $M_2=0.5$ $M_{\rm Sun}$, we get an RV amplitude of $\sim 3.9 \sin{i}$ km/s, with a period of $\sim 17$ years.

\section{Conclusion and perspectives}
This letter reports the detection of a candidate dim companion to $\epsilon$~Cephei, a $\delta$~Scuti star. This is the first image of a candidate companion, seen at 330 $\pm 50$ mas separation, only 1.1 resolution elements from the central host star ($\sim 8.6$ AU). If physically associated, the companion has a mass of $\sim 0.5$ $M_{\rm Sun}$, and a temperature of 3650 K, likely placing it at the transition between late-type K stars and early M stars. The interferometrically detected near-infrared excess is likely due to the companion, and not hot dust, while the previously reported {\it IRAS} 60 $\mu$m excess was due to source confusion. Finally, the most plausible reason why this candidate companion remained hidden so far is that {\it Hipparcos} failed to detect it because of its long period, highly eccentric orbit, while the stellar pulsations of the $\delta$~Scuti primary precluded an RV detection.


This detection was made possible by state-of-the-art high contrast imaging techniques (wavefront control and vortex coronagraphy) envisioned for upcoming next-generation ground-based extreme AO systems and future space-based coronagraphs. We have demonstrated a $5\sigma$ detection capability of the order of $\sim 2 \times 10^{-3}$ at $\sim 1\lambda/D$, which is sufficient to directly image secondary stars and brown dwarfs to an order of magnitude fainter than speckle interferometry. At $2\lambda/D$, we routinely reach $10^{-4}$ contrast levels, and with careful calibration $\sim 10^{-5}$ \citep{Mawet2010b,Serabyn2010}, two to three orders of magnitude better than current interferometric techniques. Extreme AO coupled to vector vortex coronagraphy has thus great potential for the characterization of spectroscopic binary and brown dwarf orbits, especially with a larger telescope. It can also dramatically improve imaging capabilities even with rather small telescopes. Such techniques thus open the door to the direct imaging of spectroscopic binary companions, as well as brown dwarfs and exoplanets.

\acknowledgments

This work was carried out at the European Southern Observatory (ESO) site of Vitacura (Santiago, Chile), and the Jet Propulsion Laboratory (JPL), California Institute of Technology (Caltech), under contract with the National Aeronautics and Space Administration (NASA). The data presented in this Letter are based on observations obtained at the Hale Telescope, Palomar Observatory, as part of a continuing collaboration between Caltech, NASA/JPL, and Cornell University. This work is also based (in part) on observations made with the {\it Spitzer Space Telescope}, which is operated by the JPL, Caltech, under a contract with NASA. This research has made use of the NASA/IPAC/NExScI Star and Exoplanet Database, which is operated by the JPL, Caltech, under contract with NASA, and NASA's Astrophysics Data System and of the SIMBAD database, operated at CDS (Strasbourg, France).

\end{document}